\documentstyle[epsfig]{mn}

\newcommand{\iras} {{\it IRAS \/}}

\newcommand{\be}{\begin{equation}}
\newcommand{\ee}{\end{equation}}

\newcommand{\bea}{\begin{eqnarray}}
\newcommand{\eea}{\end{eqnarray}}

\newcommand{\bc}{\begin{center}}
\newcommand{\ec}{\end{center}}

\newcommand{\lu}{\,h^{-1}{\rm Mpc}}

\renewcommand{\vec}[1]{ {\bmath #1} }

\newcommand{\dd}{{\rm d}}

\title{Estimates for the Luminosity Function of Galaxies and its Evolution}

\author[V. Springel and S. D. M. White]
{Volker Springel and Simon D. M. White \\
Max-Planck-Institut f\"{u}r Astrophysik, 
Karl-Schwarzschild-Stra\ss{}e 1, 
85740 Garching bei M\"{u}nchen, Germany}

\begin{document}

\maketitle
\begin{abstract}
A new method is presented to obtain a non-parametric maximum
likelihood estimate of the luminosity function 
and the selection function 
of a flux limited redshift survey. The method 
parameterizes the selection function as a series of step-wise
power laws and allows possible
evolution of the luminosity function. We also propose 
a new technique to
estimate the rate of evolution of the luminosity function. This is based
on a minimization of the observed large-scale power with respect to
the evolutionary model. We use an ensemble of mock surveys
extracted from a N-body simulation to verify the power of
this method.

We apply 
our estimators to
the 1.2-Jy survey of \iras galaxies. We find a far-infrared 
luminosity function 
in good agreement
with previously published results and evidence for rather strong
evolution. If the comoving 
number density of \iras galaxies is assumed to scale  $\propto (1+z)^P$, we estimate
$P=4.3\pm1.4\pm1$.

\end{abstract}

\section{Introduction}

Flux limited redshift surveys
are a major tool for studying the large-scale 
structure of the Universe. 
However, such 
catalogues show a strong decline of the mean 
number density of galaxies as a function of distance, simply because at larger
distances more and more galaxies fall below the apparent flux limit.
Most of the statistics that are used to analyse 
these surveys require an accurate knowledge of this dependence,
which is described in terms of the selection function (SF). 
Hence 
the accuracy of the adopted SF can limit the reliability of
such large-scale structure studies. 

Closely related to the SF is the luminosity function
(LF), which describes the distribution in luminosity of the galaxy
population
sampled by the particular 
redshift survey. This quantity is of more fundamental importance, since
it should be reproduced by any viable theory of
galaxy formation and evolution.

Here we revisit the problem of determining the LF and SF, given only the data
of a flux limited redshift survey. 
Current standard methods for this task 
include Schmidt's \shortcite{Sc68} $V/V_{\rm max}$  estimator, and 
the maximum likelihood techniques first introduced by 
Turner \shortcite{Tu79} and Sandage, Tamman \& Yahil \shortcite{Sa79}.
The maximum likelihood methods are 
generally superior to the older techniques because they allow the construction 
of estimators
which are not
systematically biased by density inhomogeneities. For this reason we
will focus on them in the following.

Two basic procedures 
have been used to find maximum likelihood estimates of the LF.
In the so-called parametric maximum likelihood estimate
\cite{Sa79,Ya91}
an analytic form for the LF (or SF) is assumed 
that depends on a few (typically two to
four) parameters. These are then determined by maximizing the likelihood of
the observed data set.

However, because the maximum likelihood technique offers
no built-in measure of goodness-of-fit,
almost any functional form can be made to `fit',
although the function may provide only a poor description of the data.
The parametric technique therefore requires
an {\em a priori} knowledge of a suitable fitting form.

If this information is not available one can allow a very flexible shape
of the LF by describing it by many parameters in a reasonable way. 
This so-called step-wise non-parametric maximum likelihood method
 \cite{Ni83,Ef88} has been
used  both to find luminosity
functions \cite{Ef88,Lo92,Li96} and to estimate the run of radial
density with distance \cite{Sa90,Lo92}. So far these applications only 
employed simple step functions to model the desired function.

In this paper we 
propose a non-parametric maximum
likelihood estimator that uses
a new parameterization in terms of piece-wise
power laws. 
This method provides accurate information on the shapes
of the SF and the LF and has a number of computational advantages. For example, 
it does not require iterative solutions and it provides 
error-estimates easily. The method is particularly useful for
justifying 
specific analytic
fitting forms for the SF and LF.

We also discuss ways to estimate evolution of the LF. Such evolution 
appears 
quite strong in the far-infrared. A $V/V_{\rm max}$ test 
may be used to estimate the evolutionary rate \cite{Av80}, but 
the result can be strongly affected by
density inhomogeneities. Saunders et al. \shortcite{Sa90} proposed an
alternative
 maximum
likelihood estimator for density evolution. Unfortunately, 
their technique is troubled 
by the same problem; it explicitly assumes a locally uniform distribution
of galaxies. 
In order to reduce the influence of density inhomogeneities (which are
clearly present)
we propose to combine the SF determination with a counts-in-cells analysis on large
scales. The evolutionary rate is then estimated by minimizing the variance of
the large scale density field. We demonstrate the robustness 
of this method using an ensemble of flux limited mock catalogues constructed 
by observing an N-body model
universe.

Our paper is organized as follows.
Sections 2 and 3 present in detail our non-parametric estimators for the SF
and the LF. The estimation of evolution is discussed in section 4.
Finally, we apply these methods in section 5 to the 1.2-Jy 
survey of \iras galaxies. In particular, we give results for the far-infrared
LF, its evolution and the SF of the 1.2-Jy survey.

\section{A non-parametric estimator for the selection function}

\subsection{Definitions}

Let the field $n_z(\vec{r},L)\,\dd L$ describe the comoving number density of
galaxies at epoch $z$, in the luminosity interval $[L,L+\dd L]$ and at
comoving 
spatial 
position $\vec{r}$. 
Assuming that the luminosity distribution
is independent of 
clustering the 
number
density field may be written as
\be
n_z(\vec{r},L)=\frac{n_z(\vec{r})}{\bar{n}_z} \Phi_z(L) ,
\label{A3}
\ee
where $\Phi_z(L)$ describes the LF.
Here
$n_z(\vec{r})=\int_{L_0}^{\infty} n_z(\vec{r},L) \,\dd L$ is the local number density and
$\bar{n}_z=\left< n_z(\vec{r}) \right>$ 
signifies the mean number density, averaged over many
realizations of the Universe. 
The LF is normalized as
$
\int_{L_0}^{\infty} \Phi_z(L)\,\dd L=\bar{n}_z
$
and the dependence on $z$ takes care of a 
possible time evolution, if present. The luminosity cut $L_0$ may be used
to handle a possible formal divergence of the integral over
$\Phi_z(L)$ at the lower end.

We define the selection function $S(z)$ as the mean comoving number density 
of galaxies that one expects to see in a flux limited survey at 
redshift  
$z$.
Then $S(z)$ is given
by
\be
S(z)=\int_{L_{\rm min}(z)}^{\infty} \Phi_z(L) \, \dd L \, ,
\label{A4}
\ee
where $L_{\rm min}(z)$ denotes the minimum luminosity a source at redshift $z$ 
may have without falling below the flux limit of the catalogue.
In this work we neglect the peculiar velocities of galaxies and take all
redshifts to be cosmological, i.e. we adopt a simple redshift-distance
relation $z=z(|\vec{r}|)$.

\subsection{The likelihood expression}

We now imagine that the catalogue is drawn from an underlying parent
distribution given by $n_z(\vec{r},L)$ for $L\ge L_{\rm min}(z)$ 
and by $0$
for $L < L_{\rm min}(z)$,
where $z=z(\vec{r})$.
Then the conditional probability $p(L_i|z_i)\,\dd L$ that a source observed at redshift
$z_i$ falls into the luminosity range $[L_i,L_i+\dd L]$ takes the form
\be
p(L_i|z_i)\,\dd L=\frac{n_{z_i}(\vec{r_i},L_i)\,\dd L}{\int_{L_{\rm
min}(z_i)}^{\infty}n_{z_i}(\vec{r_i},L')\,\dd L'}
\, .\label{EQ77}
\ee
The denominator simply counts the available number of galaxies at that distance
and the numerator gives the number of galaxies in the particular luminosity
range.
Upon insertion of equations (\ref{A3}) and (\ref{A4}) this becomes
\be
p(L_i|z_i)=\frac{\Phi_{z_i}(L_i)}{S(z_i)}.
\label{A6}
\label{EQ3}
\ee  
Note that the density fluctuations have dropped out of this
expression and so the dependence on $\vec{r}$ can be dropped in
equation (\ref{EQ77}).
This insensitivity to density inhomogeneities
makes the maximum likelihood technique used here
superior compared to older methods like the ordinary $V/V_{\rm max}$
estimator.
If one now maximizes the likelihood
\be
{\cal L}=\prod_i p(L_i|z_i)
\label{EQ2}
\ee
of the whole data set with respect to $S(z)$ one obtains an estimate of the
SF that is not systematically biased by local density
fluctuations.

In order to find this maximum in practice we first express
$\Phi_{z_i}(L_i)$ in terms of the SF with the 
help of equation (\ref{A4}).
Here a model 
for the evolution of the LF has to
be specified. For brevity we will only treat a case with pure density
evolution according to
\be
\Phi_z(L)=g(z)\Phi_0(L).
\ee
However, our method can be easily
generalized to include luminosity evolution as well.

We further define a maximal redshift $z_i^{\rm m}$ for each source such that 
$
L_{\rm min}(z_i^{\rm m})=L_i 
$.
If then the derivative $S'(z)$ of equation (\ref{A4}) 
is evaluated at $z_i^{\rm m}$ one 
obtains
\be
S'(z_i^{\rm m})=-\frac{g(z_i^{\rm m})}{g(z_i)} \Phi_{z_i}(L_i) L_{\rm
min}'(z_i^{\rm m})
+ \frac{g'(z_i^{\rm m})}{g(z_i^{\rm m})}\: 
S(z_i^{\rm m})\, ,
\ee
so that the probability of equation (\ref{A6}) can be expressed entirely in
terms of $S(z)$ and $g(z)$.
Hence one finally has to maximize 
\bea
\Lambda=\ln{\cal L} & = &
\sum_i \ln \left(
\frac{-S'(z_i^{\rm m})}{S(z_i)}+ \frac{g'(z_i^{\rm m})}{g(z_i^{\rm m})}\: 
\frac{S(z_i^{\rm m})}{S(z_i)}
\right) \nonumber\\
& &+\sum_i \ln\left( \frac{g(z_i)}{g(z_i^{\rm m})} \right) 
-\sum_i \ln L'_{\rm min}(z_i^{\rm m}),
\label{EQ1}
\eea
where the constant sum of the last term 
may be dropped.

The above form suggests that one might be able to 
maximize $\Lambda$ simultaneously
for $g(z)$ and $S(z)$. However, if 
equation (\ref{A6}) is rewritten for the density evolution model it becomes
\[
p(L_i|z_i)=\frac{g(z_i)\Phi_0(L_i)}{\int_{L_{\rm
min}(z_i)}^{\infty}g(z_i)\Phi_0(L)\,\dd L}=
\frac{\Phi_0(L_i)}{\int_{L_{\rm min}(z_i)}^{\infty}\Phi_0(L)\,\dd L}.
\]
So the function $g(z)$ drops out completely and there is no sensitivity
to density evolution with this estimator.
In fact, this was to be expected since 
estimates based on the likelihood (\ref{A6}) are
independent of the density distribution by construction.

\subsection{A non-parametric maximum likelihood estimator}

Here we propose a 
new variant of the non-parametric
method that models $S(z)$ as a series of continuously linked power laws.
This description seems 
appropriate since the SF is a smooth curve that covers a wide
range of values and its local behaviour can be very well approximated
by a power law.

We describe $S(z)$ by $n$ pieces. Let $S_k=S(x_k)$ be
the values of $S(z)$ at a series of ascending redshifts $x_k$ where $k\in\{1,2,\ldots,n\}$. Then bin
$1$ covers $0<z\leq x_1$, bin $2$ covers $x_1<z\leq x_2$, and so forth.
In each piece $k$, $S(z)$ is taken to be a power law of the form
\be
S(z)=S_k \left( \frac{z}{x_k} \right)^{m_k} \hspace{1cm} \mbox{for
$x_{k-1}< z \leq x_k$}\, ,
\ee
where $m_k$ is the logarithmic slope of the particular piece. 
These slopes are precisely the
quantities needed to characterize the shape of the SF.

The different pieces have to join continuously, because 
 $S(z)$ is an integral over the LF. 
Continuity requires the
$S_k$ to be related by
\be
S_k=
S_1 \displaystyle \prod_{j=2}^k
\left(\frac{x_j}{x_{j-1}}\right)^{m_j} \hspace{0.6cm}   \mbox{for
 $1<k\le n$ .}
\label{A7}
\ee

Let us
further define $a_i$ as the number of the bin to which the maximal redshift
$z_i^{\rm m}$ of galaxy $i$ belongs. Similarly, let $b_i$ be the number of the
interval
that encloses the 
redshift $z_i$ of galaxy $i$.
Then the likelihood (\ref{EQ1}) takes the form  
\bea
\Lambda&=&\sum_i\left[ \ln\left( \frac{-m_{a_i}}{z_i^{\rm
m}}+\frac{g'(z_i^{\rm m})}{g(z_i^{\rm m})} \right)
+\sum_{j=b_i+1}^{a_i}m_j\ln \frac{x_j}{x_{j-1}}\right] \nonumber \\
& &
+\sum_i m_{a_i} \ln\frac{z_i^{\rm m}}{x_{a_i}}
-\sum_i m_{b_i} \ln\frac{z_i}{x_{b_i}} \nonumber \\
& &
+ \sum_i \ln\frac{g(z_i)}{g(z_i^{\rm m})}
-\sum_i\ln {L'_{\rm min}(z_i^{\rm m})}.
\eea
The best estimates for the $m_k$ can be found by solving the
likelihood equations
\be
\frac{\partial \Lambda}{\partial m_k} =
\sum_i\frac{\delta_{k,a_i}}{m_k-z_i^{\rm m }\frac{g'(z_i^{\rm m})}{g(z_i^{\rm m})}}+T_k =0,
\label{A8}
\ee
where $T_k$ is defined as
\bea
T_k &=& 
\sum_i \sum_{j=b_i+1}^{a_i}\delta_{j,k}\ln \frac{x_j}{x_{j-1}} 
+\sum_i \delta_{k,a_i} \ln\frac{z_i^{\rm m}}{x_{a_i}} \nonumber \\
& & -\sum_i \delta_{k,b_i} \ln\frac{z_i}{x_{b_i}} .
\label{A10}
\eea

The equations (\ref{A8}) are not fully linear in $m_k$, but {\em almost}. Since the
redshift bins are narrow, replacing $z_i^{\rm m}$ by $x_{a_i}$ will give a useful
first approximation $\tilde{m}_k$ to the true solution $m_k$, which might then
quickly be improved by an iteration technique. This
starting value can be calculated as
\be
\tilde{m}_k=-\frac{n_k}{T_k}+x_k\frac{g'(x_k)}{g(x_k)} ,
\label{A9}
\ee
where $n_k$ is the number of galaxies in bin $k$. Note that in the
case of no evolution
the solution to equation (\ref{A8}) is simply given by $m_k=-{n_k}{T_k}^{-1}$.

The maximum likelihood method also allows an estimate of the
statistical uncertainties of the derived parameters \cite{Ke79}.
Asymptotically the
distribution of $\cal L$ is a multivariate Gaussian around the true values of
the parameters. 
If the information matrix $I$ is defined as
\be
I_{ij}=-\frac{\partial^2\Lambda}{\partial m_i \partial m_j},
\ee
then the covariance matrix of the parameter estimates is given by
${\rm cov}(m_i,m_j)=(I^{-1})_{ij}$
evaluated at the maximum of $\Lambda$.
This result holds asymptotically for large sample sizes.

Because the $T_k$ don't explicitly depend on any of the $m_k$ the
information matrix is simply found to be
\be
I_{lk}=\delta_{l,k}
\sum_i 
\frac{\delta_{k,a_i}}
{\left(m_k -z_i^{\rm m}\frac{g'(z_i^{\rm m})}{g(z_i^{\rm m})}\right)^2}.
\ee
One can therefore trivially invert this diagonal matrix and find error
estimates for the $m_k$ as 
\be
{\rm var}(m_k)=\left[\sum_i 
\delta_{k,a_i}
{\left(m_k -z_i^{\rm m}\frac{g'(z_i^{\rm m})}{g(z_i^{\rm m})}\right)^{-2}}
\right]^{-1}\simeq
\frac{m_k^2}{n_k}.
\ee

This is a surprisingly simple result. In
particular, one can solve for each of the $m_k$ independent of all the
others. The $m_k$ are mutually uncorrelated which shows that it is
essentially the
local logarithmic slope of the SF
that is determined by the maximum likelihood estimator (\ref{EQ2}). 

Once the non-parametric estimate of the shape of the SF is found,
it can be used to find and justify an appropriate analytic fitting
function for $S(z)$. 
For this purpose one can directly employ a minimum $\chi^2$ fit of the
$m_k$ to the logarithmic slope of some fitting form for $S(z)$. 
Of course, once an analytic 
form has been selected in this way, the values of its
parameters can also be determined with the parametric maximum
likelihood technique by using the fitting form directly in equation (\ref{EQ1}).

\subsection{Normalization}

Because the normalization of the SF is lost with the above estimator
it has to be found in a second step. For this purpose 
we write the SF as 
\be
S(z)=\psi s(z),
\ee
where $s(z)$ is the shape of the SF as determined above.
Then an unbiased estimate
$\tilde{\psi}$ of the factor $\psi$ is given by 
\be
\tilde{\psi}=\frac{\int_V m(\vec{r})w(\vec{r})\,\dd\vec{r}}{\int_V
s(\vec{r}')w(\vec{r}')\,\dd \vec{r}'}
\label{EQ444}
\ee
for an arbitrary weight function $w(\vec{r})$. 
Here 
$m(\vec{r})=\sum_i \delta(\vec{r}-\vec{r}_i)$
represents
the observed galaxy field and we employ the 
shorthand notation $s(\vec{r})=s(z(|\vec{r}|))$.

Following Davis \& Huchra \shortcite{Da82} we choose the weight function
$w(\vec{r})$ 
such that the expected variance of $\tilde{\psi}$ is minimized.
This is to a good approximation the case for 
\be
w(\vec{r})=\frac{1}{1+J_3\tilde{\psi}s(\vec{r})},
\label{A7A}
\ee
where
$
J_3=\int_V 4\pi r^2 \xi(r) \,\dd r
$
is the second moment of the two-point correlation function and $V$
denotes the volume used in the normalization.

Note that the normalization estimate has to be found
iteratively since the estimator depends implicitly on
$\tilde{\psi}$. 
One also needs a model for the two-point
correlation function in order to estimate $J_3$. However, this is
uncritical because the
dependence on $J_3$ is usually quite weak. Hence an estimate 
of $J_3$ based on a linear theory CDM power spectrum should be
entirely sufficient.

The numerical value of $\tilde{\psi}$ itself does not provide
a meaningful measure for the local number density of galaxies. In
fact, such a measure is not well defined for a purely flux limited
catalogue. Instead, the normalization is perhaps best expressed in
terms of the expected number density $N$ of galaxies per unit solid angle
on the sky, i.e.
\be
N=\frac{1}{4\pi}\int S(z) \frac{{\rm d}V}{{\rm d}z}{\rm d}z.
\ee

\section{A non-parametric estimator for the luminosity function}

Using equation (\ref{A4}) 
it is possible to recover the underlying luminosity
distribution 
if the SF is known, or vice versa. 
In particular one can readily derive a non-parametric
LF estimate from the SF estimator described
in the previous section.

We suppose that the SF has been determined with the 
estimator described above and that the function
$L_{\rm min}(z)$ and its inverse
$z_{\rm max}(L)$ are known.
The piecewise description of the SF directly
translates into a piecewise description of the LF
if boundaries $L_k$ of luminosity intervals are defined by
$L_k\equiv L_{\rm min}(x_k)$.
Upon evaluation of the derivative of equation (\ref{A4}) at $z=z_{\rm max}(L)$ the present
day LF results as
\bea
\Phi_0(L) & =&   \left( \frac{g'(z_{\rm max})}{g(z_{\rm max})}S(z_{\rm max})-
S'(z_{\rm max}) \right) \nonumber\\
& &\times
 \left( {g(z_{\rm max})L'_{\rm min}(z_{\rm max})} \right)^{-1}.
\eea
This translates into the piecewise description
\bea
\Phi_0(L) & = & \left(\frac{g'(z_{\rm max})}{g(z_{\rm max})}-\frac{m_k}{z_{\rm max}} \right)
\left( \frac{z_{\rm max}}{x_k} \right)^{m_k} \nonumber \\
& & \times\frac{S_k}{g(z_{\rm max}) L'_{\rm min}(z_{\rm max})} 
\label{A19}
\eea
for $L_{k-1}<L<L_k$, which is the desired non-parametric estimate.

The usual way to find a 
non-parametric maximum
likelihood estimate of the LF utilizes a
parameterization by a series of step
functions \cite{Ef88,Lo92}. 
We see two main advantages of our method as compared to this approach.

First, the shape of the LF over each bin 
is approximately a power
law, so it is able to adapt to the true
shape of  the
LF in a flexible way. Only very small discontinuities at
the boundaries of bins remain and the estimate $\Phi_0(L)$ of
equation (\ref{A19}) traces the smooth LF quite well. 
The conventional estimator on
the other hand assumes constant $\Phi_0$ over each bin which leads 
to large
discontinuous jumps in the LF at bin boundaries. 
Efstathiou et~al. \shortcite{Ef88}
show that the heights $\Phi_k$ of these bins are related by $\Phi_k
\approx \int_{L_k-\Delta L/2}^{L_k+\Delta L/2}\, \Phi \,\dd N(L) /  \int_{L_k-\Delta L/2}^{L_k+\Delta L/2}
\,\dd N(L)$
to the underlying luminosity distribution. It is therefore more
complicated to infer the smooth underlying LF from
this non-parametric estimate.

A second nice feature is that the absence of correlations among the $m_k$
simplifies the estimation of errors in the LF.
To demonstrate this let us define the values
$
\Phi_k \equiv \Phi_0(L_k)
$
of the LF at the points $L_k$. In appendix A we compute the covariance
matrix 
\be
V_{kl}={\rm cov}(\ln \Phi_k , \ln \Phi_l)
\ee
of these quantities. The matrix $V_{kl}$ may then 
be used to fit an analytic model
$\tilde{\Phi}_k$ to the non-parametric LF estimate by minimizing the covariance form
\be
\chi^2=
\sum_{l,k}(\ln\Phi_l-\ln\tilde{\Phi}_l)V^{-1}_{lk}(\ln\Phi_k-\ln\tilde{\Phi}_k).
\label{EQCF}
\ee

Although not of major importance a further practical 
advantage of the estimator presented here
is its relative computational
ease. In particular one does not have to resort to lengthy iteration
techniques as required by the conventional parameterization with a series of
step functions.

\section{Estimating evolution}

\subsection{The radial density distribution}

As was demonstrated above, the estimators based on equation (\ref{EQ3})
are insensitive
to density evolution. If the latter 
is important (as appears to be the case in the \iras surveys) 
a different approach is needed 
to determine its rate.

Instead of $p(L_i|r_i)$ as before we can equally well write down 
the conditional probability
\be
p(r_i|L_i)\,\dd r=\frac{4\pi r_i^2 n_{z_i}(r_i,L_i) \,\dd
r}{\int_0^{r_i^{\star}} n_{z_i}(r,L_i) \,\dd V}
=\frac{4\pi r_i^2 n_{z_i}(r_i) \,\dd r}{\int_0^{r_i^{\star}} n_z(r) \,\dd V}
\label{A3334}
\ee
to find a galaxy with luminosity
$L_i$ 
in a distance interval of width
$\dd r$ at coordinate $r_i<R_{\rm max}$. 
Here the density field has been averaged over direction 
and the definition
$r_i^{\star}\equiv\min(r_i^{\rm m},R_{\rm max})$ allows an upper
distance cut-off $R_{\rm max}$ to be included. We consider the
inclusion of the latter
to be important because the density field is only known well
to a finite depth.  
At large
distances it is dominated by shot noise due to
increasingly sparse sampling. In addition the completeness 
of most surveys is worst in this regime.

Equation (\ref{A3334}) shows that the density
distribution $n_z(r)$ can be
estimated independent of the LF if 
\be 
\Lambda=\sum_i \ln p(r_i|L_i)=\sum_i \ln \frac{n_{z_i}(r_i)}{\int_0^{r_i^{\star}}
n_{z(r)}(r) \,\dd V}
\label{A14}
\ee
is maximized with respect to $n_z(r)$.  
Saunders et al. \shortcite{Sa90} have used a parameterization of $n_z(r)$ in terms of a
series of step functions leading to a non-parametric radial density
estimator. This estimator is quite useful since it can indicate
the presence of evolution and it can show the most prominent density
features. 

\subsection{The constant density method \label{Sec1}}

In order to obtain an estimate of the evolutionary rate 
Saunders et~al. \shortcite{Sa90} proposed replacing 
$n_z(r)$ in equation (\ref{A3334}) by
the expected mean density $\bar{n}_{z}=\bar{n}_0g(z)$ at epoch
$z=z(r)$.
Adopting an evolutionary model, for example
\be
g(z)=(1+z)^P ,
\ee
one can then find the rate by maximizing
\be
\Lambda=\sum_i \ln \frac { (1+z_i)^P}{ \int_{z_{\rm min}}^{z_i^{\star}}
(1+z)^P \frac{\dd V}{\dd z} \,\dd z}\, 
\label{A16}
\ee
with respect to $P$. A number of subsequent studies \cite{Fi92,Ol94}
also applied this method to \iras galaxies.

Of course, in reality the galaxies are not distributed uniformly. This
fact leads to a serious drawback of the above estimator; the resulting 
estimate of $P$ can be strongly influenced by density inhomogeneities.
For example, 
an overdensity in the foreground mimics negative evolution, whereas an
overdensity in
 the background can bias the evolution low.

Since this effect should be strongest nearby, 
a lower redshift cut-off $z_{\rm min}$, as introduced above,
may be used 
to
partly reduce its influence.
However, as will be seen in our application to the 1.2-Jy survey the
evolutionary estimates can depend quite strongly on the particular 
redshift interval chosen and it is unclear which choice gives the most reliable
answer.

\subsection{The minimal variance method\label{SecV}}

We here want to propose an 
alternative evolutionary estimator that 
avoids the unphysical neglect of density fluctuations and that 
gets rid of the
arbitrary choice of redshift interval used in equation
(\ref{A16}). 

As shown above  
the data of the redshift survey alone do not allow us to disentangle 
density evolution and large-scale structure. This is only possible if
the fundamental assumption of statistical homogeneity 
is added. In a pragmatic approach we might then base the determination
of evolution on the notion
that the correct rate should lead to 
a universe that looks as close as possible to homogeneous on very large
scales. As a quantitative measure for deviations from homogeneity one might,
for example, 
take the variance $\sigma^2$ of the density field smoothed on some large
scale $\lambda$. 
The determination of the evolution might
therefore be formulated in terms of an extremal principle:
The evolutionary rate can be estimated by minimizing the observed 
power on large
scales. 

To estimate $\sigma^2$ we apply a slightly modified version of the method of 
Saunders et~al. \shortcite{Sa91}. 
We first compute a smoothed galaxy density field 
\be
d_i=\sum_j w_{ij}\frac{m_j}{S_j}
\label{EQ8}
\ee
on a fine mesh. Here $m_{j}$ denotes the number of galaxies in cell $j$, $S_j$ is
the value of the SF and $w_{ij}=W(\vec{r}_i-\vec{r}_j)$ a smoothing kernel, which we
here take to be a Gaussian of the form
\be
W(\vec{x})= \frac{1}{\pi^{3/2}\lambda^3}
\exp\left(-\frac{ |\vec{x}|^2}{\lambda^2} \right).
\ee

Then unbiased estimators for the mean $\bar{d}$ and the variance $\sigma^2$ 
of the galaxy density field 
are given by
\be
\bar{d}=\frac{\sum_i g_i d_i}{\sum_i g_i}
\ee
and
\be
\sigma^2=\frac{\sum_i h_i \left[ \left(\frac{d_i-\bar{d}}{\bar{d}}\right)^2 - Y^{(2)}_i
\right]}
{\sum_i h_i},
\ee 
where
\be
Y^{(n)}_i= \bar{d}^{\, 1-n} \sum_j \frac{w_{ij}^n}{S_j^{n-1}}.
\ee
These estimators become minimum variance estimators in a good approximation 
if
the weights are chosen as 
\be
g_i=\left[{Y_i^{(2)}}\right]^{-1}
\ee
and
\bea
h_i & = &\left[
Y^{(4)}_i
+2 \left(Y^{(2)}_i\right)^2 + 2 \sigma^4 \right. \nonumber \\
& & \left.
\; +\left(3\left(Y^{(2)}_i\right)^2
+4Y^{(3)}_i
+4Y^{(2)}_i\right) \sigma^2 
\right]^{-1}
\label{Y4}.
\eea

Since the estimate of the variance depends on the SF it
becomes a function of the adopted evolutionary model used in
computing the SF. It is therefore possible to estimate the evolutionary rate by
minimizing $\sigma^2$.

A difficulty with this technique is that it does not easily supply an error
estimate for the evolutionary rate. Additionally it is unclear how to choose $\lambda$ in an 
optimal way. A large $\lambda$ allows the survey volume to be probed 
to greater depth,
thereby increasing the sensitivity to evolution. However, the number of
independent smoothing volumes declines simultaneously, thereby degrading
the accuracy of the variance estimate. What is then the 
optimal choice?

We solve both of these problems by extracting an ensemble of mock catalogues
from a large N-body simulation. The mock surveys mimic the statistical
properties of the survey at hand in terms of the SF,
the evolution,
the sky coverage and the source density.
This suite of catalogues can be used to find the optimal
$\lambda$ and to estimate the precision of the final evolutionary estimate.

Although computationally somewhat lengthy this scheme is well worth
the effort because it offers 
an increased accuracy as we will demonstrate with the
mock surveys in our application to the 1.2-Jy survey.
A further advantage is that it
is equally applicable to luminosity evolution, in contrast to the
estimator discussed in section \ref{Sec1}.

\section{Application to the 1.2-Jy Redshift Survey}

The data of the 1.2-Jy redshift survey \cite{St90} of \iras galaxies has been
published \cite{Fi95} and can be retrieved electronically from the
Astronomical Data Center (ftp://adc.gsfc.nasa.gov).
The 5321 galaxies of the survey are selected from the PSC catalogue above a
flux limit of 1.2$\,$Jy in the 60$\,\mu$m band. The sky coverage is 87.6
per cent,
excluding only the zone of avoidance for $|b|<5^{\rm o}$ and 
a few unobserved or contaminated patches at higher latitude.  

We convert the redshifts to
the Local Group frame and use them subsequently to infer distances without
further corrections for peculiar velocities.

\subsection{K-correction and maximal redshift}

The estimators presented above demand an accurate computation of the maximal 
redshift 
$z_i^{\rm m}= z_{\rm max}(L_i)$ that a given source
could have
without falling below the flux limit. A detailed calculation of this quantity
involves K- and
colour-corrections, which require a knowledge of the spectral energy
distribution (SED) of the source, a specification of the responsivity 
$R(\nu)$ of
the detector, and a choice for the cosmological
background model.

For the case of the \iras surveys Fisher et~al. \shortcite{Fi92} 
have phrased this problem 
in terms of the useful equation
\be
\frac{f_{\rm min}}{f_{i}} =
\frac{r_i^2}{r(z_i^{\rm m})^2}\:\frac{\Psi(\eta)}{\eta},
\label{eq4}
\label{A22}
\ee
which implicitly specifies $z_i^{\rm m}$ in a general form. 
Here $f_i$ is the quoted flux density of a source and $r_i$ is its
comoving coordinate
distance. $f_{\rm min}$ gives the flux limit of the catalogue and 
$\eta$ is defined as $\eta=(1+z_i^{\rm m})(1+z)^{-1}$.
The function
\be
\Psi(\eta)=
\frac{\int R(\nu) f_{\nu}\left(\nu\eta\right)\,\dd\nu}
{\int R(\nu') f_{\nu}(\nu')\,\dd\nu'} \label{psi}
\ee
encodes K- and colour-corrections
by means of 
the observed flux density $f_{\nu}(\nu)$ 
of the source.
The cosmological background model enters via the function $r=r(z)$. We
will adopt an Einstein-de-Sitter universe, i.e.
\be
r(z)=\frac{2}{a_0H_0}\left(1-\frac{1}{\sqrt{1+z}}\right).
\ee

The 60$\,\mu$m band SED of \iras galaxies 
might be approximated as a straight
power law $L_{\nu}(\nu)\propto \nu^\alpha$ with $\alpha\approx -2$. In
this case $\Psi(\eta)=\eta^\alpha$.
We have typically used $\alpha=-2$, but 
also tried a gray-body model \cite{Sa90} fitted to $60\,\mu $m/$100\,\mu$m
flux ratios and the polynomial model of Fisher et~al. \shortcite{Fi92}
which results in a considerably shallower SED.

\subsection{Selection function}

In our determination of the SF and LF we will assume that
the LF exhibits density evolution of the form $g(z)=(1+z)^P$ with
$P=4.3$. The actual estimation of the evolutionary rate is discussed
below in section \ref{SecEvol}.

\begin{figure}
\bc
\resizebox{8cm}{!}{\includegraphics{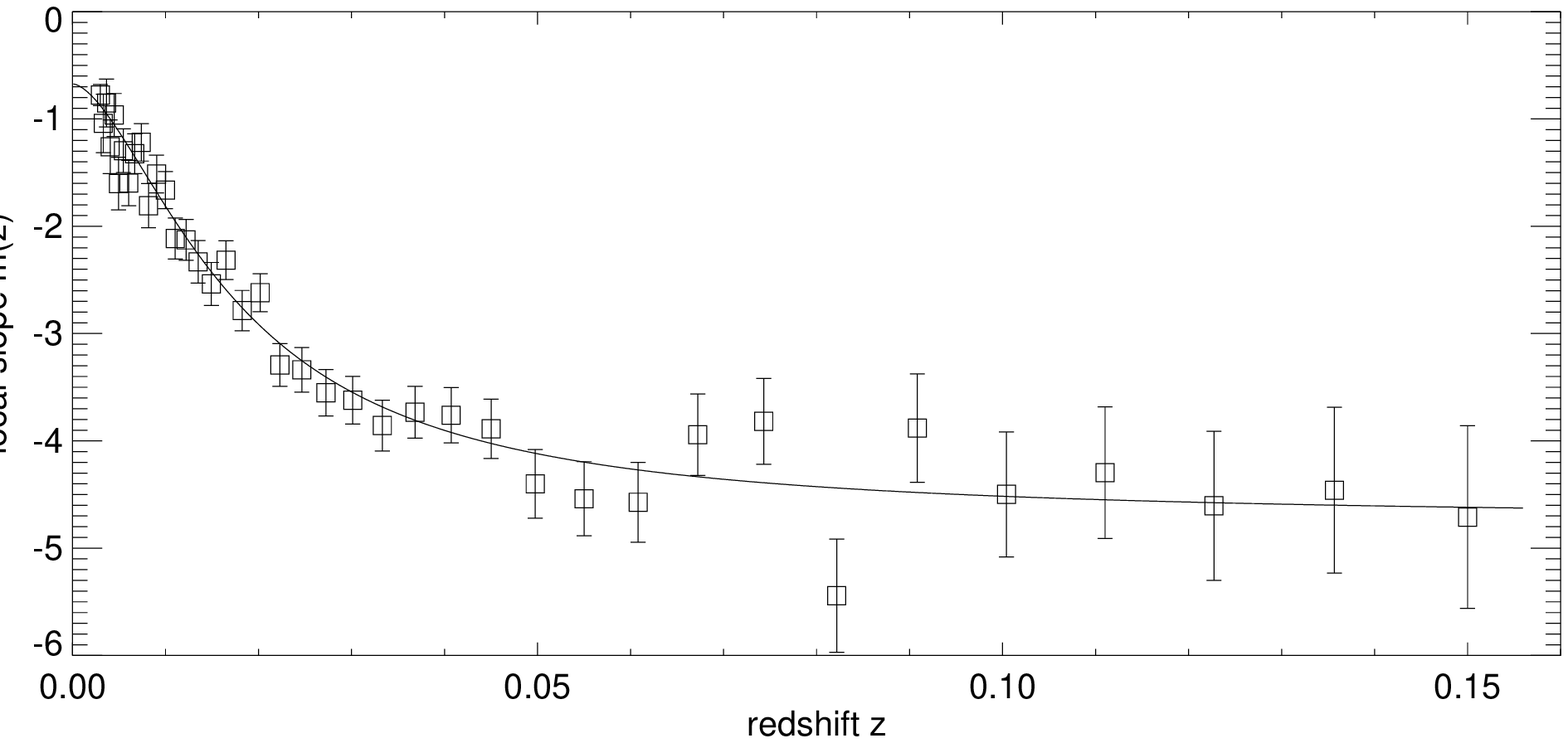}}
\resizebox{8cm}{!}{\includegraphics{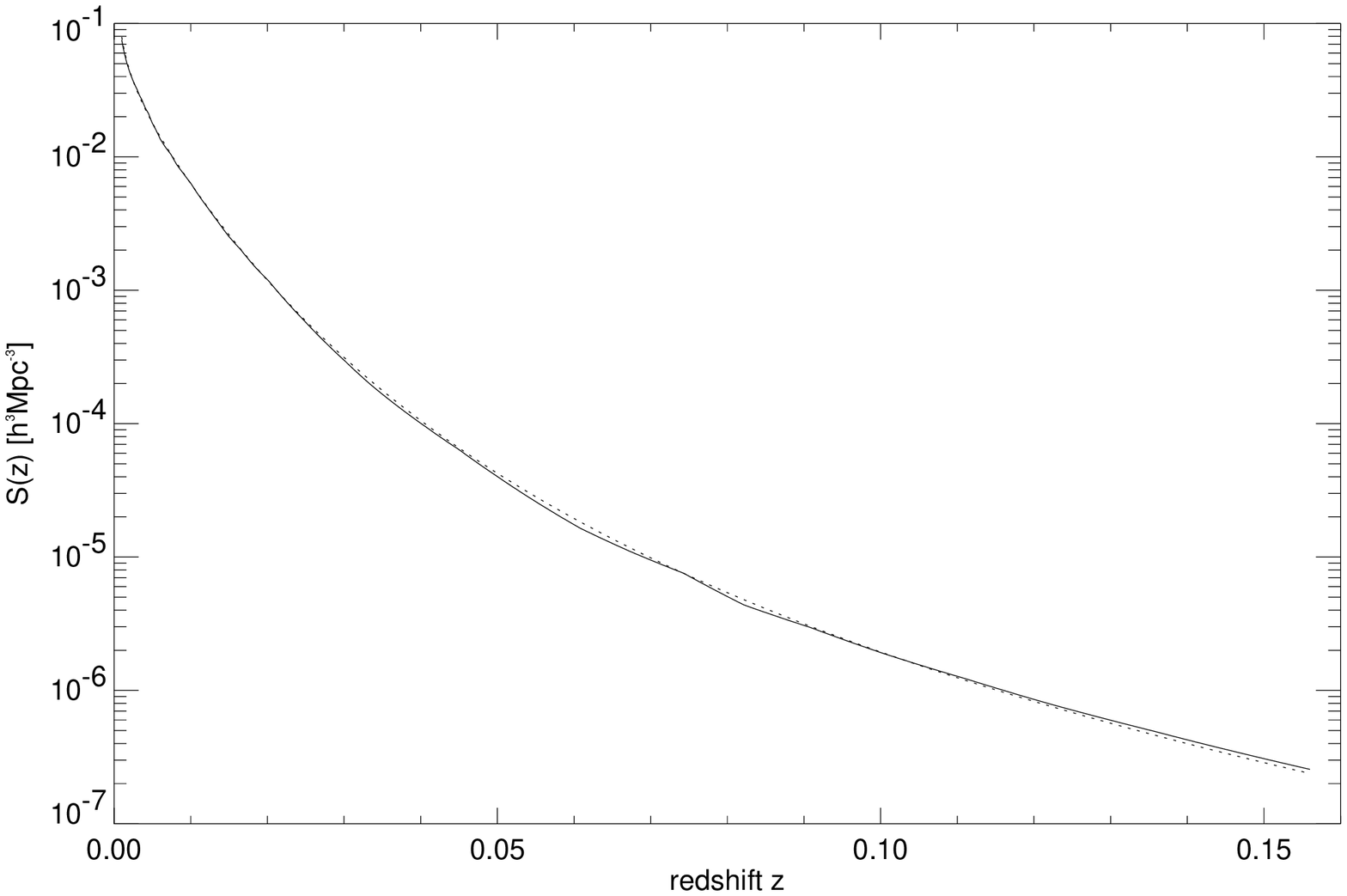}}
\ec
\caption[Selfunc]
{Non-parametric selection function estimate. The top panel
shows the estimated slopes $m_k$ with $1\sigma$ error bars. 
The fit is  based on equation (\ref{EQ10}).
The lower panel compares the actual non-parametric SF (solid) and the
analytic fit (dotted) with the parameters of table 1.
\label{Fig1} }
\end{figure}

Figure \ref{Fig1} shows the estimated local slopes $m_k$ of the SF and a fit based on the functional form
\be
S(z)=\frac{\psi}{z^{\alpha}\left(
1+\left(\frac{z}{z^{\star}}\right)^{\gamma}\right)^{\beta/\gamma}}.
\label{EQ10}
\ee
We used 40 redshift intervals, logarithmically spaced between
$x_0=0.003$
and $x_{39}=0.15$, for the non-parametric estimate. 
A minimum $\chi^2$ fit of the measured $m_k$ to the form of
equation (\ref{EQ10}) resulted in a reduced $\chi^2_\nu$ of 0.87 
(for $\nu=36$ degrees of freedom),
which indicates a good fit. This form of the SF is slightly more
general than the one used by Yahil et~al. \shortcite{Ya91} and Fisher
et~al. \shortcite{Fi95} who
essentially fixed $\gamma$ at the value 2. This gives a marginally
worse  $\chi^2_\nu$ of 0.91.

Our best estimates for the final SF parameters are listed in table
\ref{tab1}. They are obtained by maximizing the likelihood of the
parametric form directly. Although a fit to the non-parametric estimate gives a
very close result, we prefer these numbers as final estimates, because
they are free of any binning effects. For each of the 
parameters $\alpha$, $\beta$,
$\gamma$, and $z^\star$ the quoted errors give 
$1\sigma$ confidence intervals obtained from the bounding box of the 
$\Lambda_{\rm max}-0.5$ likelihood contour.

\begin{table}
\label{tab1}
\bc
\caption{Parameters of the selection function fit.}
\begin{tabular}{ccc}
$\alpha$  & $\beta$ & $\gamma$ \\
$0.741^{+0.128}_{-0.135}$  
&  $4.210^{+0.419}_{-0.344}$ 
& $1.582^{+0.237}_{-0.214}$ \\
\\
$z^{\star}$ & $\psi \;\;[h^3{\rm Mpc}^{-3}]$ & \\
$0.0184^{+0.00213}_{-0.00167}$ 
&$(486.5\pm 13.0 ) \times 10^{-6}$ & \\
\end{tabular}
\ec
\end{table}

\begin{figure}
\bc
\resizebox{8cm}{!}{\includegraphics{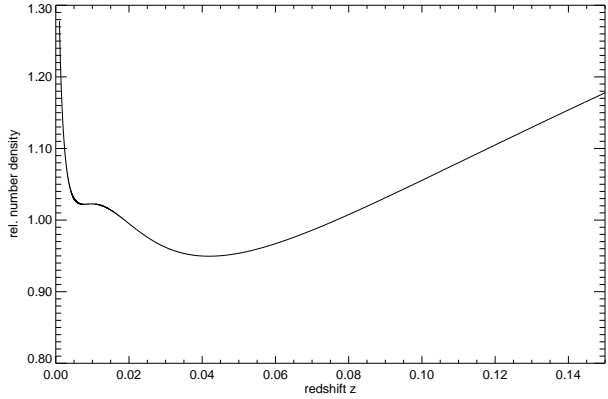}}
\ec
\caption[Selfit]
{Ratio of the 1.2-Jy-SF of Fisher et~al. \protect\shortcite{Fi95}
to the result of this
work. The ratio is scaled such that both parameterizations of the SF
predict the same number of galaxies out to redshift 0.15. \label{FigRatio}
}
\end{figure}

Figure \ref{FigRatio} shows the ratio of Fisher et~al.'s
\shortcite{Fi95} SF to our estimate. The agreement is quite good out to
redshift around 0.1. However, relative to our result the predicted mean
density of
galaxies at redshift 0.15 is about 20 per cent higher in the Fisher
et~al. \shortcite{Fi95} result.

In order to normalize the SF we used the volume inside $z=0.15$. This
determines the parameter $\psi$ 
and results in 
$N  =  (490\pm 13)\;{\rm sr}^{-1}$ expected galaxies per unit
solid angle on the sky. Our final LF is also normalized to this
number. 
The $1\sigma$ error-estimate 
is based on a direct computation
of the variance of $\psi$ by means of equation (\ref{EQ444}). Here the
uncertainty in the shape of $s(z)$ is neglected.

\subsection{Luminosity function}

\begin{figure}
\bc
\resizebox{8cm}{!}{\includegraphics{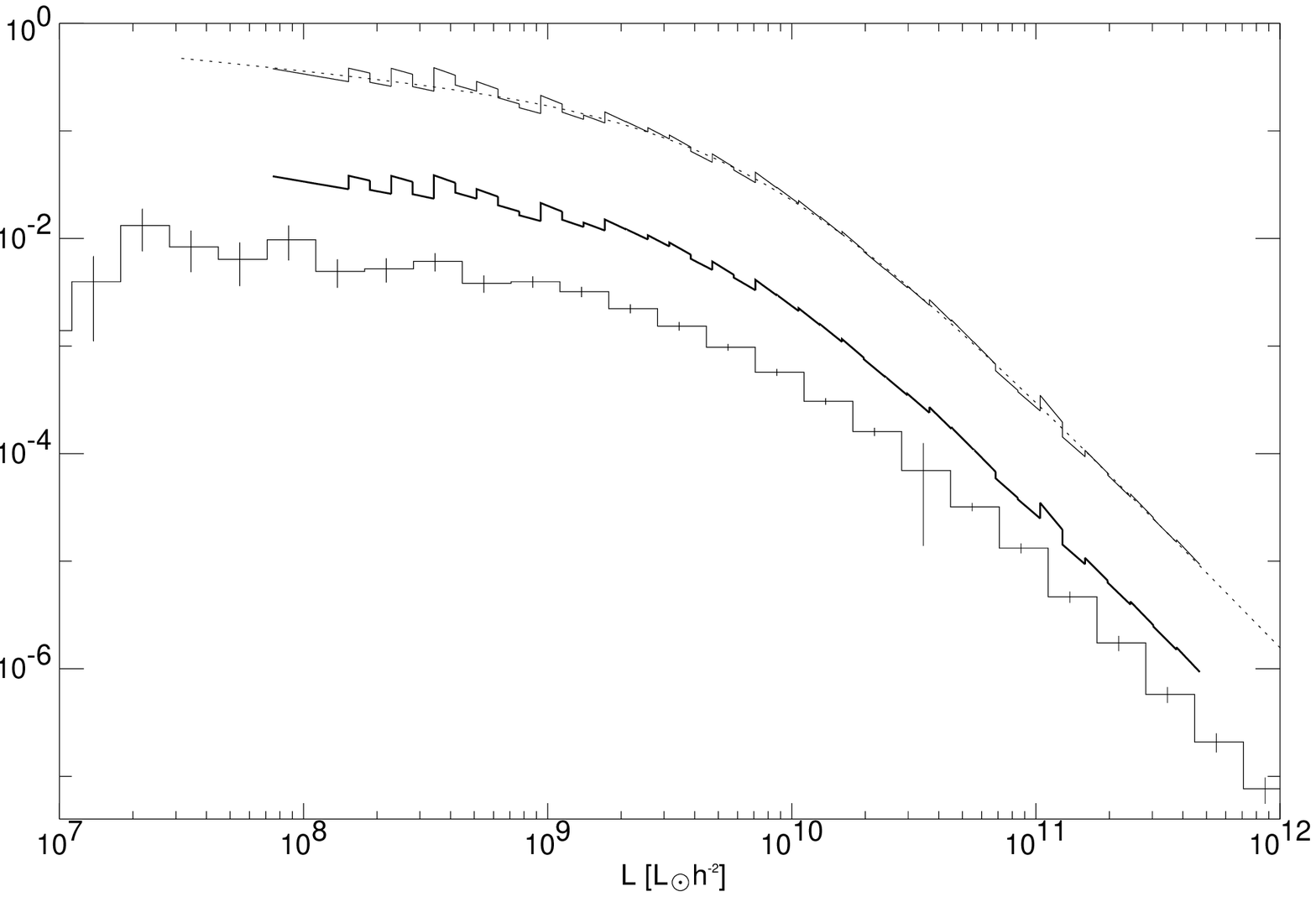}}
\resizebox{8cm}{!}{\includegraphics{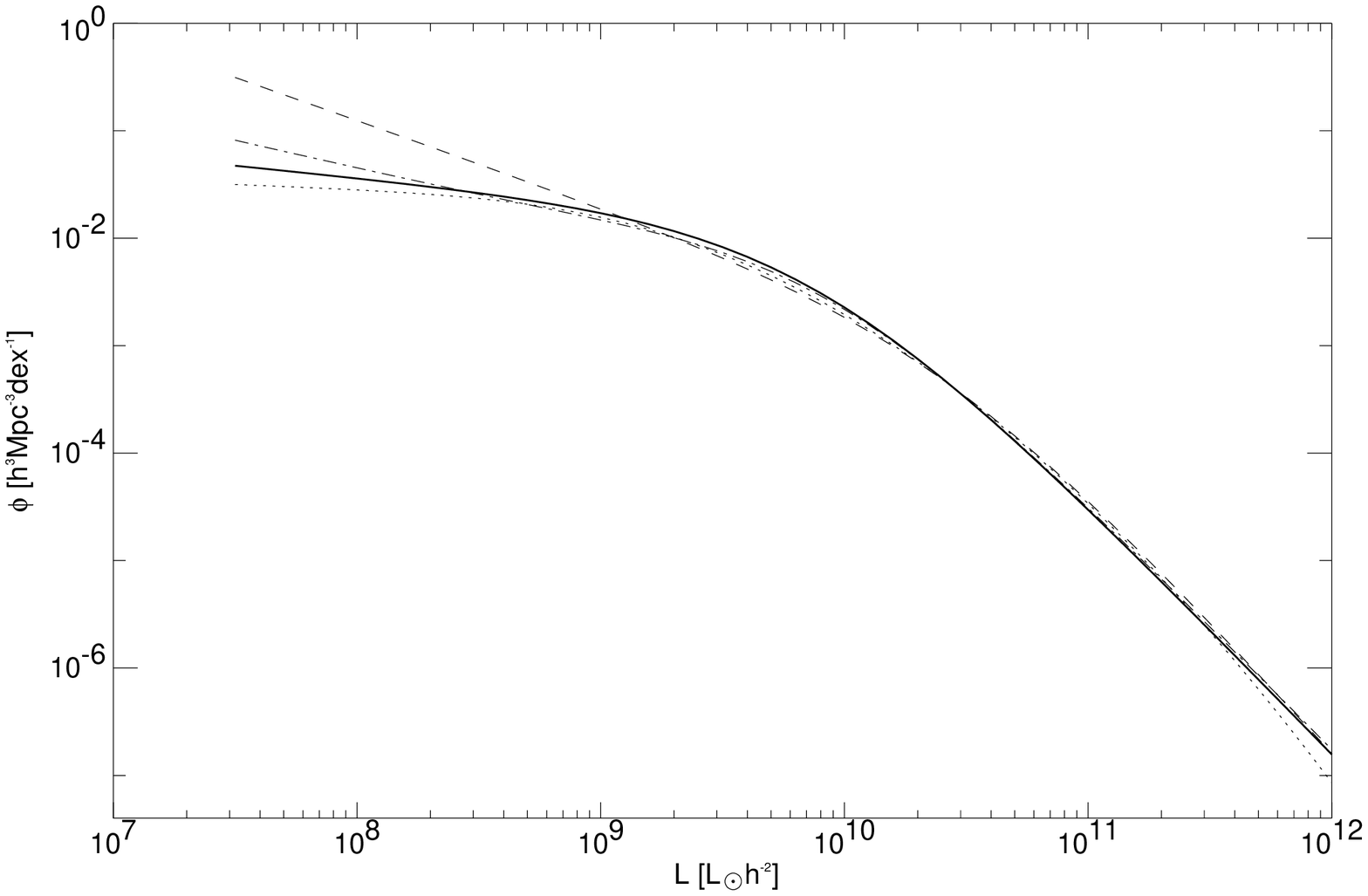}}
\ec
\caption[Luminosity]
{Luminosity function estimates. 
The top panel compares our non-parametric LF estimate (thick)
with the estimate of Saunders et~al. \protect\shortcite{Sa90}. The
latter has been rescaled to $H_0=100 h {\rm km}\,{\rm s}^{-1}$ and
shifted vertically for graphical clarity. The two upper curves (also
shifted) compare the non-parametric estimate with our analytic fit.

The lower panel shows various parameterization of the far-infrared LF
by different authors: Saunders et~al. \shortcite{Sa90} (dotted), Yahil
et~al. \shortcite{Ya91} (dot-dashed), 
Lawrence et~al. \shortcite{La86} (dashed), this work (thick).
\label{FigLum} }
\end{figure}

Figure \ref{FigLum} displays our non-parametric estimate of the far-infrared LF
resulting from the method outlined in section 3. For comparison also
shown is the non-parametric estimate of Saunders et~al. \shortcite{Sa90}
which takes
the form of a histogram because it uses step functions to model the
LF. 
In order to facilitate a comparison with the literature 
we present the LF in terms of luminosity per
decimal decade of luminosity, viz.
\be
\phi(L)=\Phi_0(L)\,L\,\ln 10,
\ee
and we define $L$ for a source with SED $L_\nu(\nu)$ as
$L=\nu\,L_\nu(\nu)$ at 60$\,\mu$m.

\begin{figure}
\bc
\resizebox{8cm}{!}{\includegraphics{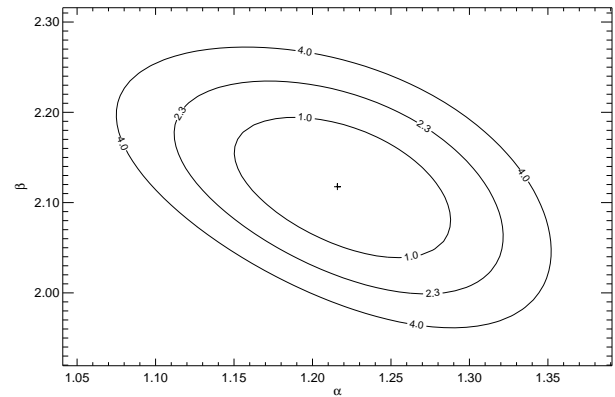}}
\ec
\caption[Vontour]
{Contour levels of constant $\Delta\chi^2$ for the LF fit 
in the subspace of $\alpha$ and
$\beta$. The thick contour defined by $\Delta\chi^2=2.3$ encloses the
68.3 per cent joint confidence region of $\alpha$ and $\beta$. 
\label{FigCont} }
\end{figure}

We believe that our parameterization offers an improved description of
the LF, without large unphysical jumps due to
binning of sparse data. In particular, the quality of a fit with an
analytic function can be judged quite easily.

As a fitting form 
we use the function
\be
\phi(L)=C \left(\frac{L}{L_{\star}}\right)^{1-\alpha}
\left( 1+\frac{L}{L_{\star}\beta}\right)^{-\beta}
\label{Aj}
\ee
proposed by Lawrence et~al. \shortcite{La86} and give the best-fit 
parameters in table \ref{TabLum}. The fit is based on a minimization
of the covariance form (\ref{EQCF}), imposing a fixed
normalization as described in appendix A. 
The cited errors for $\alpha$,
$\beta$ and $L_{\star}$ 
give 68.3 per cent confidence intervals derived from an
increase of $\Delta\chi^2=\chi^2-\chi^2_{\rm min}$ by 1, where
$\Delta\chi^2$ is marginalized in the space of $\alpha$, $\beta$, $L_{\star}$, and $C$.
Figure \ref{FigCont} shows
contour levels of constant $\Delta\chi^2$ in
the subspace of $\alpha$ and $\beta$. The contour of
$\Delta\chi^2=2.3$ encloses the 68.3 per cent joint confidence region of
$\alpha$ and $\beta$.
The error given in table \ref{TabLum} for the normalization $C$ is
computed 
for a fixed shape of the
LF.

The simple-two power law (\ref{Aj}) does a remarkably
good job in fitting the measurements 
as evidenced by the comparison in figure \ref{FigLum} and by
a reduced $\chi_\nu^2$ of $1.06$.
In particular we
find no need to choose a functional form
that shows more curvature
over the full range of luminosities \cite{Sa90}.
However, we must not forget that
we have ignored velocity fields in this paper and that we cannot go as
faint as Saunders et~al. \shortcite{Sa90} who used an additional set
of very local galaxies.
This might affect the
faint end slope; we find a value not quite as shallow as Saunders
et~al. \shortcite{Sa90}, but somewhat flatter than Yahil
et~al. \shortcite{Ya91}.

\begin{table}
\bc
\caption{Parameters of the luminosity function fit.}
\label{TabLum}
\begin{tabular}{cc}
$\alpha$  & $\beta$  \\
$1.221^{+0.068}_{-0.072}$ 
& $2.116^{+0.080}_{-0.079}$ \\
\\
$L_{\star} \;\;[h^{-2}L_{\odot}]$ & $C \;\;[h^3{\rm Mpc}^{-3}]$ \\
$3.615^{+0.640}_{-0.568}\times 10^9$
& $(1.670 \pm0.045)\times 10^{-2}$ \\
\end{tabular}
\ec
\end{table}

\subsection{Evolution \label{SecEvol}}

Evidence for evolution in \iras galaxies has been reported by a number
of authors using differential source counts
\cite{Ha91,Ha87b,Lo90,Gr95} and redshift surveys \cite{Lo89,Sa90,Fi92,Ol94}.
However, there has been some
controversy about the magnitude of the evolutionary rate seen in the
far-infrared LF.
Saunders et~al. \shortcite{Sa90} found $P=7\pm 2$ for the QCD survey, whereas
Fisher et~al. \shortcite{Fi92} gave $P=2 \pm 3$ for the 1.2-Jy survey, a result
apparently consistent 
with no evolution.
Oliver et~al. \shortcite{Ol94} found $P=6.25\pm 1.5$ 
for a survey of
faint \iras galaxies and $P=4.2\pm 2.3$ for the QDOT survey. 
More recently Bertin et~al. \shortcite{Be96} report $P=6.0\pm 1.2$ 
for the FIR sample.

The low result for the 1.2-Jy survey, obtained with the method of section 
\ref{Sec1}, 
was based on an early version of
the catalogue. Since evolutionary estimates are quite sensitive to
completeness we here analyse the final catalogue again. This is also necessary
because the evolutionary rate is needed as input parameter
for the determination of the SF and LF.

\begin{figure}
\bc
\resizebox{8cm}{!}{\includegraphics{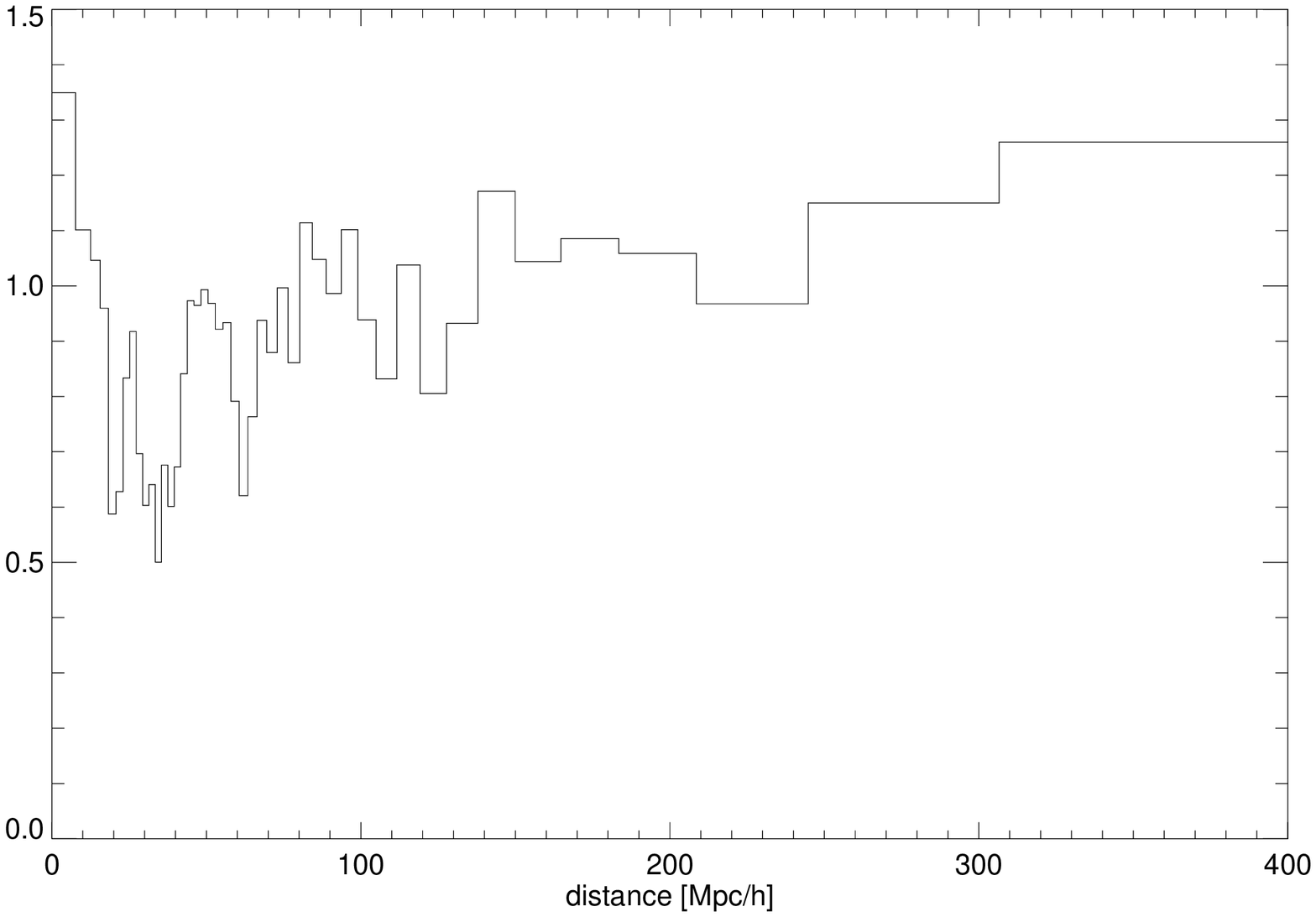}}

\resizebox{8cm}{!}{\includegraphics{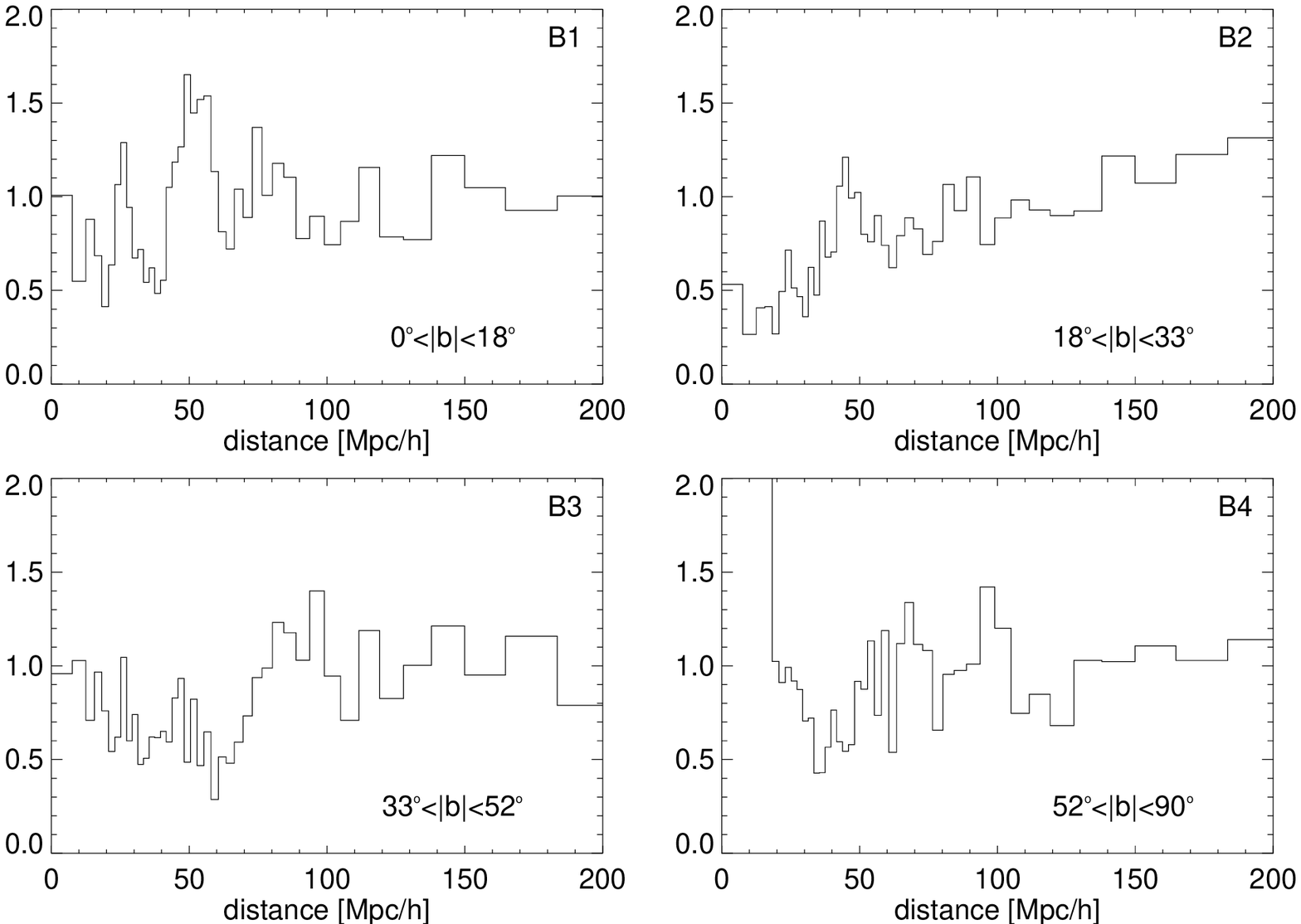}}
\ec
\caption
{Radial density distribution obtained with a non-parametric
estimator based on equation (\ref{A14}) and the assumption of no evolution.
The result for the full sample is shown in the top panel, whereas the
plots labeled B1--B4 are computed for subsamples selected by galactic
latitude: $|b|<18^{\rm o}$ (B1), $18^{\rm
o}\le |b|<33^{\rm o}$ (B2), $33^{\rm o} \le |b|<52^{\rm o}$ (B3), and
$52^{\rm o} \le |b|$ (B4).
\label{FigDens}}
\end{figure}

\subsubsection{The run of radial density}

A first indication that a correct treatment of evolution is important
may be obtained with the 
radial density estimator based on equation (\ref{A14}). 
We adopt a parameterization of
$n_z(r)$ in terms of 45 step-functions (with $R_{\rm max}=400\lu$)
of widths chosen such that the 
signal-to-noise ratio in each bin stays roughly constant. 

In the resulting distribution, displayed in figure \ref{FigDens}, 
the average
density seems to rise with distance. 
A similar behaviour is also found
if smaller patches of the sky are examined separately. 
As an illustration we have split the sample in 4 regions of
approximately equal size selected in terms of the modulus $|b|$ of the
galactic latitude. In this way we also obtain a 
low latitude sample which might 
be
less complete than the other parts of the survey. In detail these
regions, labeled B1-B4, cover $|b|<18^{\rm o}$ (B1), $18^{\rm
o}\le |b|<33^{\rm o}$ (B2),
$33^{\rm o} \le |b|<52^{\rm o}$ (B3), and $52^{\rm o} \le |b|$ (B4).
As evidenced by figure \ref{FigDens} the rise of density with distance 
can be seen in all four
subsamples, albeit with variable strength. Prominent structures are also
visible, in particular Perseus-Pisces (around $50\lu$) and the Local
Supercluster, which is responsible for the nearby overdensity seen in B4.

\begin{table}
\caption{Evolutionary estimates $P$ obtained with the constant density
method for various redshift intervals. The subsamples B1--B4 are
selected in terms of galactic latitude: $|b|<18^{\rm o}$ (B1), $18^{\rm
o}\le |b|<33^{\rm o}$ (B2), $33^{\rm o} \le |b|<52^{\rm o}$ (B3), and $52^{\rm o} \le |b|$ (B4).
\label{TabP}}
\bc
\begin{tabular}{lrrrr}
 &  $[0,\infty]$  & $[0.01,\infty]$ & $[0,0.1]$ & $[0.01,0.1]$\vspace{0.1cm}\\
Full& $3.7\pm 1.5$ & $5.0\pm 1.6$ & $3.9\pm 1.9$ & $5.9\pm 2.0$ \\ 
B1 & $-1.0\pm3.3$    & $-1.5\pm3.4$ & $2.6\pm3.9$  & $2.5\pm4.2$ \\
B2 & $9.2\pm3.2$    & $6.1\pm3.2$   & $11.1\pm3.8$ & $6.7\pm3.9$ \\
B3 & $6.1\pm3.1$    & $7.8\pm3.3$   & $7.1\pm3.7$  & $9.7\pm4.0$ \\
B4 & $0.6\pm2.8$    & $6.5\pm3.1$   & $-6.2\pm3.8$ & $4.1\pm4.3$ \\
\end{tabular}
\ec
\end{table}

\subsubsection{Evolutionary estimates with the constant density method}

We now turn to estimates of the density evolution parameter $P$ 
obtained with the estimator of section \ref{Sec1}.
Table \ref{TabP} shows results 
for various redshift
intervals, in each case computed 
for the full catalogue and for the subsamples B1--B4.

Although the statistical errors are large, they are insufficient 
to explain
the large differences in the estimates that result when the redshift
boundaries are changed.
For example, when
the upper redshift cut-off is lowered from $\infty$ to $0.1$, the
low latitude sample B1 shows a marked increase in $P$ which is likely to be the
result of incompleteness at large $z$. The sample B4 on the other
hand exhibits a very strong overdensity nearby, that 
biases evolution low if it is not excluded by imposing a lower redshift
cut-off.

We have tried many different redshift intervals and find it
hard to come to definite conclusions with this method, simply because the
result is strongly affected by density inhomogeneities and the 
choice of a redshift interval exhibits a certain degree of arbitrariness.
We therefore  
prefer to estimate evolution with the minimum variance method outlined
in section \ref{SecV}.

\subsubsection{An ensemble of mock surveys}

In order to demonstrate that this method is indeed 
more reliable we have constructed 
an ensemble
of 1.2-Jy-like mock catalogues 
by {\em observing} a large N-body simulation of a
standard CDM universe. The simulation has been kindly provided by the
Virgo
collaboration \cite{Je96}. It
contains $256^3$ particles in a periodic box of size $V=L^3$ with $L=240\lu$ 
and is normalized to fluctuations of $\sigma_8=0.6$ in spheres of
$8\lu$. The details of this model are not of much relevance for the
following, the salient point is that the model exhibits clustering in realistic
strength.

By randomly placing an observer in the periodic box 
we constructed 50 mock catalogues of this simulation. We used 
a
SF very similar to the 1.2-Jy survey (with parameters $\alpha= 0.84$,
$\beta=3.96$, $\gamma=1.74$, $z^{\star}=0.018$) and imposed density evolution with $P=5$. 
The
mock catalogues featured the same angular mask as the 1.2-Jy survey 
and were sampled to a density of 5321 galaxies inside $460 \lu$.
The suite of these catalogues 
simulates Poisson
sampling noise and -- to a reasonable degree -- cosmic variance as
well. 
The catalogues
can be ideally used to test the performance of the
various estimators discussed in this paper.

First we applied the constant density estimator of section \ref{Sec1} to these
catalogues. Using the redshift interval $[0,0.1]$ we obtained 
$\left<P\right>=5.7$ with an rms scatter of 2.4 among the
individual measurements. This scatter is somewhat larger than the
statistical error of $\approx 1.9$ provided by the likelihood method
itself. Smaller redshift intervals
also led to estimates $\left<P\right>$ close to 5, however with larger scatter.

Next we applied our new minimum variance method to the mock catalogues. For
this purpose we computed maximum likelihood SF fits 
for 11 different values of $P$ in the range 0
to 10 individually for each catalogue. This allows the computation of
a curve $\sigma^2_\lambda(P)$ for each catalogue, and -- by fitting
the minimum with a parabola -- an estimate of $P$.

We treated the technical difficulty of the presence of an angular mask
in the surveys by using the ratio method of Melott \& Dominik
\shortcite{Me93} in the smoothing process. This means that we actually
chose
\be
w_{ij}=\frac{W(\vec{r}_i-\vec{r}_j)}
{\int W(\vec{r}_i-\vec{r}') M(\vec{r}')
\,\dd\vec{r}'}
\ee
as kernel in equation (\ref{EQ8}), where $M(\vec{r})$ is a field equal to 1 in
the actual survey volume and equal to 0 in the volume behind the
angular mask. The estimation of the  variance is only done with cells
inside the survey volume.

The method finally requires the choice of a smoothing length $\lambda$. 
Since the minimum variance
estimator automatically turns off when the signal-to-noise ratio
becomes too low a larger $\lambda$ usually means that the  
the survey is probed to higher depth, thereby
increasing the sensitivity to evolution. On the other hand, the accuracy of 
the measurement of the variance degrades with increasing $\lambda$
because the number of independent smoothing volumes in the usable
survey volume declines. We find a 
choice for $\lambda$ that represents a compromise between
these two effects by minimizing the rms scatter of the
$P$ measurements for the mock catalogues.

For this purpose we 
have computed estimates of $\left< P\right>$ and its scatter for a
range of smoothing lengths and find that the uncertainty is
smallest for $\lambda \simeq 60 \lu$, however, any $\lambda$ in the range
50-80$\lu$ works almost equally well.
For $\lambda=60 \lu$
we obtain $\left< P\right>
= 5.3$ with rms fluctuations of 1.4. This is a considerable improvement
compared to the constant density method, nearly cutting the random
error in half. This is possible because the new method eliminates most of the
systematic influence of density inhomogeneities on the estimate of
$P$. The remaining uncertainty is mainly due to counting statistics.

\subsubsection{Evolutionary estimate with the minimum variance method}
 
The above results show that the new 
estimator for density evolution gives a more precise estimate than
the constant density method. Hence we apply it now to obtain
our best estimate for the density evolutionary rate of the 1.2-Jy
survey.
Adopting $\lambda= 60 \lu$ and using the same procedure as applied to the
mock surveys we find $P=4.31\pm 1.4$, where the error estimate is based on the
mock surveys.

Clearly, this error estimate is somewhat too optimistic, because the
1.2-Jy survey is not as perfectly selected 
as the mock catalogues. Additionally there are a number of
systematic uncertainties that degrade the accuracy of the evolutionary
measurement. 

For example, the SED might be shallower as we assumed here, leading to
an overestimate of evolution. The gray-body model (we find $P=3.9$) does only
very little change compared to the straight $\alpha=-2$ SED used here.
On the other hand, Fisher et~al.'s \shortcite{Fi92} significantly flatter 
polynomial model lowers the evolutionary rate quite strongly to
$P=3.0$. However, the SED of this model is likely to be too
flat. In addition to $60\,\mu$m and $100\,\mu$m  
it tries to simultaneously fit the $25\,\mu$m fluxes which are
typically underpredicted by the gray-body model. However, only
half of the galaxies (the bright ones at $25\,\mu$m) 
have detections in this band at all, so the polynomial model is biased towards a
shallow SED.  
The polynomial model will also underpredict the slope
of the SED if the spectra are indeed comprised of a cool and
a warm component, as seems often to be the case \cite{Ro89}.

Because the 1.2-Jy survey is quite local, a
change of background cosmology to a low density universe has 
very little effect, increasing the estimated $P$ very slightly.

More serious are potential problems with the \iras flux scale. 
Either a baseline flux error with accompanying
Malmquist bias, or a nonlinear flux scale could lead to a significant
overestimate of the evolutionary rate. 
However, so far there is no convincing evidence that the \iras PSC
is troubled with flux errors of the strength required
to explain a significant fraction of the evolutionary signal.
For example, Oliver et~al. \shortcite{Ol94} give an upper limit of
36$\,$mJy for the baseline flux error of the PSC. At this level,
Malmquist bias is not an issue, as we have checked with 
mock surveys that exhibit artificial flux errors.

We believe that our new method for estimating the evolutionary rate has
removed most of the uncertainty due to density
inhomogeneities. However, sample incompleteness can bias the evolution
low if a fraction of the faint high redshift objects is missed. At low
latitude, this 
is indeed quite likely the case for the 1.2-Jy survey.

We think that an error of $\pm1$ represents a reasonable 
estimate of these systematic uncertainties.
Our final estimate for
evolution is therefore
$P=4.3\pm 1.4\pm 1$.
Given the uncertainties, this result is in good
agreement with other
determinations of $P$ cited above. However, 
is is notably higher
than Fisher et al.'s \shortcite{Fi92} 
result for an early version of the 1.2-Jy
survey.

\section{Discussion}

An accurate determination of the SF of a redshift
survey is a prerequisite for taking full advantage of its information about
the large-scale structure of the Universe. For example, statistical
methods that rely on estimates of the density field in large survey 
volumes, like power spectrum measurements or genus statistics, depend
crucially on a precise knowledge of the mean density as a function of
distance.

In this work we have
proposed a flexible non-parametric maximum likelihood 
estimator for the SF and LF. 
The method is independent of
density inhomogeneities, gives accurate information on the shapes of
SF and LF, and provides estimates of the statistical uncertainties of
the derived quantities with relative
computational ease. 
We think that the technique should be useful for upcoming redshift
surveys.

We have also proposed a new method to estimate evolution of the LF,
based on the notion that the Universe should look homogeneous on
large scales. With an ensemble of mock surveys we have demonstrated
that the uncertainties due to density inhomogeneities, which troubled
previous estimators, can be greatly reduced in this way.

In our application of these estimators to the 1.2-Jy survey of \iras
galaxies we have found evidence for strong evolution, confirming 
reports by several previous authors. 
Expressed in terms of density evolution $\propto(1+z)^P$, we find
$P=4.3\pm1.4\pm 1$.
This high evolutionary rate for the far-infrared LF
is hard to explain as a statistical fluke. 
If confirmed the strong evolution of
the far-infrared LF will represent
an interesting challenge for theories of
galaxy formation, interaction and evolution.

\bibliography{procdbl}

\appendix
\section{Covariance matrix of the non-parametric luminosity function estimate}

In this section we compute the covariance matrix 
\be
V_{kl}={\rm cov}(\ln \Phi_k , \ln \Phi_l)
\ee
of the non-parametric LF estimate. Because we are primarily interested
in the uncertainty of the shape of the LF we impose 
a fixed normalization, i.e. the
quantity
\be
N' =  \int_0^{x_n}S(z)z^2 \dd z  =  S_1 Q
\ee
is held constant. Here $Q=Q(m_1,\ldots,m_n)$ is given by
\be
Q=\sum_{k=1}^{n}\frac{x_k^3}{m_k+3}\left[1-\left(\frac{x_{k-1}}{x_k}\right)^{m_k+3}\right]
\prod_{j=2}^k \left(\frac{x_j}{x_{j-1}}\right)^{m_j}.
\ee
Then $\ln\Phi_k$ may be expressed as
\bea
\ln\Phi_k & = &
\ln\left(\frac{g'(x_k)}{g(x_k)}-\frac{m_k}{x_k}\right)+\sum_{j=2}^{k}m_j
\ln\frac{x_j}{x_{j-1}}
\nonumber \\
& & - \ln Q + \ln\frac{N'}{g(x_k)L'_{\min}(x_k)}.
\eea
An expansion of $\ln\Phi_k$ to linear order in the slopes $m_i$ allows
an estimate of the covariance matrix in the form
\be
V_{kl}=\sum_{i=1}^{n}A_{ki}A_{li} {\rm var}(m_i),
\ee
where we have defined 
\be
A_{ki}=\frac{\partial \ln\Phi_k}{\partial\, m_i}.
\ee
The expressions for $A_{ki}$ are somewhat lengthy, yet straightforward
to calculate:
\bea
A_{ki} & = & \frac{\delta_{ki}}{m_k-x_k\frac{g'(x_k)}{g(x_k)}}
+\sum_{j=2}^{k}\delta_{ij}\ln\frac{x_i}{x_{i-1}} \\
& & -\frac{1}{Q}\left\{ U_i -   \left.\frac{S_i}{S_1}
\frac{x_i^3}{(m_i+3)^2} \right.\right. \nonumber \\
& & \left.
\times\left[1-
\left(\frac{x_{i-1}}{x_i}\right)^{m_i+3}\left(1-(m_i+3)\ln\frac{x_{i-1}}{x_i}\right)\right]
\right\}. \nonumber 
\eea
Here $U_i$ is given by
\be
U_i=\ln\frac{x_i}{x_{i-1}}\sum_{l=i}^n
\frac{S_l}{S_1}\frac{x_l^3}{m_l+3}\left[1-\left(\frac{x_{l-1}}{x_l}\right)^{m_l+3}\right]
\ee
for $i>1$ and by $U_i=0$ for $i=1$.

\end{document}